\documentclass[showpacs]{revtex4}
\usepackage{epsfig}
\def\beq{\begin{equation}}
\def\enq{\end{equation}}
\def\beqa{\begin{eqnarray}}
\def\enqa{\end{eqnarray}}
\def\MeV{\nobreak\,\mbox{MeV}}
\def\GeV{\nobreak\,\mbox{GeV}}

\def\pli{p^\prime}

\def\ka{\kappa}
\def\la{\lambda}

\def\si{\sigma}

\def\lb{\label}
\def\nn{\nonumber}
\newcommand{\rag}{\rangle}
\newcommand{\lag}{\langle}

\begin{document}

\title{\sc QCD sum rule approach for the scalar mesons as four-quark states}
\author{Thiago V. Brito, Fernando S. Navarra, Marina Nielsen}
\affiliation{Instituto de F\'{\i}sica, Universidade de S\~{a}o Paulo, 
C.P. 66318, 05389-970 S\~{a}o Paulo, SP, Brazil.}
\author{Mirian E. Bracco}
\affiliation{Instituto de F\'{\i}sica, Universidade do Estado do Rio de 
Janeiro, 
Rua S\~ao Francisco Xavier 524, 20550-900 Rio de Janeiro, RJ, Brazil}
\begin{abstract}
We study the two point-function for the scalar mesons $\sigma,~\kappa,~
f_0(980)$ and $a_0(980)$ as diquak-antidiquark states. We also study
the decays of these mesons into $\pi\pi$, $K\pi$ and $K\bar{K}$. We found that
the couplings are
consistent with existing experimental data, pointing in favor 
of the four-quark structure for the light scalar mesons.
\end{abstract}

\maketitle

%\section{Introduction}

It is known that the identification of scalar
mesons is difficult experimentally, and that the underlying structure
of them is not well stablished theoretically,
due to the complications of the nonperturbative strong interactions. 
Actually, the observed light scalar states below $1.5\GeV$
are too numerous 
\cite{PDG} to be accommodated in a single $q\bar{q}$ multiplet.

The experimental 
proliferation of light scalar mesons is consistent with two nonets, one below
1 GeV region  and another one near 1.5 GeV. If the light scalars (the
isoscalars $\si(500),\;f_0(980)$, the isodoublet $\kappa$ and
the isovector $a_0(980)$) form an SU(3) flavor nonet, in the naive quark
model the flavor structure of these scalars would be:
\beqa
\si={1\over\sqrt{2}}(u\bar{u}+d\bar{d}),\;\;\;\;\;\;&&f_0=s\bar{s},
\nn\\
a_0^0={1\over\sqrt{2}}(u\bar{u}-d\bar{d}),\;\;\;\;\;\;&&a_0^+=u\bar{d},
\;\;\;\;\;\;a_0^-=d\bar{u},
\nn\\
\ka^+=u\bar{s},\;\;\;\;\;\;\ka^0=d\bar{s},\;\;\;\;\;\;&&\bar{\ka}^0=s\bar{d},
\;\;\;\;\;\;\ka^-=s\bar{u}.
\enqa
Although with this model it is difficult to understand the mass
degeneracy of $f_0(980)$ and $a_0(980)$, and it is hard to explain
why $\si$ and $\ka$ are broader than $f_0(980)$ and $a_0(980)$, its use is 
not yet discarded \cite{torn,bev,sca,faz,phi}. Some alternative models allow
a mixing between the isoscalars. However, different experimental data
lead to different mixing angle \cite{cheng,bnn,gsy}.

By the other hand, the scalar mesons in the $1.3 ~-~1.7\GeV$ mass region
(the
isoscalars $f_0(1370)),\;f_0(1500)$, the isodoublet $K_0^*(1430)$ and
the isovector $a_0(1450)$) may be easily accommodated in an SU(3) flavor nonet.
Therefore, theory and data are now converging that QCD forces are at work 
but with different dynamics dominating below and above 1 GeV mass.
Below 1 GeV
the phenomena point clearly towards an $S-$wave attraction among two
quarks and two anti-quarks, while above 1 GeV it is the $P-$wave $q\bar{q}$
that is manifested. \cite{cloto}.

Below 1 GeV the inverted structure of the four-quark dynamics in $S$-wave
is revealed with $f_0(980),\,a_0(980),\,\kappa$ and $\sigma$ 
symbolically given by \cite{jaffe}
\beqa
\si=ud\bar{u}\bar{d},\;\;\;\;\;\;&&f_0={1\over\sqrt{2}}(us\bar{u}\bar{s}+
ds\bar{d}\bar{s}),
\nn\\
a_0^-=ds\bar{u}\bar{s},\;\;\;\;\;\;&&a_0^0={1\over\sqrt{2}}(us\bar{u}\bar{s}
-ds\bar{d}\bar{s}),\;\;\;\;\;\;a_0^+=us\bar{d}\bar{s},
\nn\\
\ka^+=ud\bar{d}\bar{s},\;\;\;\;\;\;&&\ka^0=ud\bar{u}\bar{s},\;\;\;\;\;\;
\bar{\ka}^0=us\bar{u}\bar{d},
\;\;\;\;\;\;\ka^-=ds\bar{u}\bar{d}.
\enqa
This is supported by a recent lattice calculation \cite{lat}. In this 
four-quark scenario for the light scalars, the mass degeneracy of 
$f_0(980)$ and $a_0(980)$ is natural, and the mass hierarchy pattern of 
the nonet is understandable. Besides, it is easy to explain why $\si$ and
$\ka$ are broader than $f_0$ and $a_0$. The decays $\si\to\pi\pi$,
$\ka\to K\pi$ and $f_0,~a_0\to KK$ are OZI superallowed without the need 
of any gluon exchange, while $f_0\to\pi\pi$ and $a_0\to\eta\pi$ are OZI
allowed as it is mediated by one gluon exchange. Since $f_0(980)$ and 
$a_0(980)$ are very close to the $\bar{K}K$ threshold, the $f_0(980)$
is dominated by the $\pi\pi$ state and $a_0(980)$ is governed by
the $\eta\pi$ state. Consequently, their widths are narrower than $\si$
and $\ka$.

In the four-quark scenario it is also easier to understand why,
in some three-body decays of charmed mesons, the intermediate light scalar
meson accounts for the main contribution to the total decay rate. For
example, about half of the total decay rate of $D^+\to\pi^-\pi^+\pi^+$,
$D^+\to K^-\pi^+\pi^+$ and $D_s^+\to\pi^-\pi^+\pi^+$  comes from
$D^+\to\sigma\pi^+$, $D^+\to\kappa\pi^+$ and $D_s^+\to f_0(980)\pi^+$
repectively \cite{E791},
while the light scalar mesons are hardly seen in the semileptonic decays
$D^+\to\pi^-\pi^+\ell^+\nu_\ell$,
$D^+\to K^-\pi^+\ell^+\nu_\ell$ and $D_s^+\to\pi^-\pi^+\ell^+\nu_\ell$
\cite{ig}.

\begin{figure}[h] 
\label{fig1}
%\leavevmode
\centerline{\psfig{figure=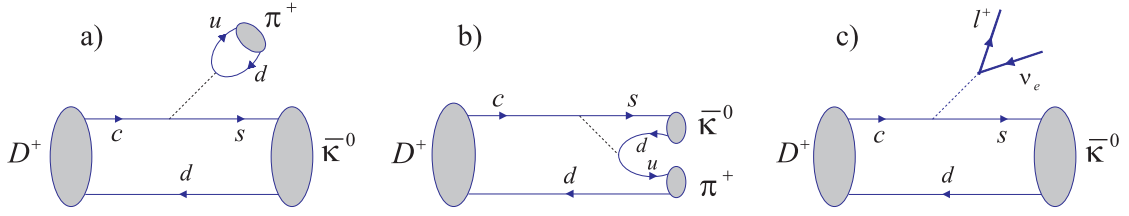,width=15cm,angle=0}}
\caption{The $D^+\to\bar{\ka}^0$ transition where $\bar{\ka}^0$ is described
as a $s\bar{d}$ state. a) and b) nonleptonic decay; c) semileptonic decay.}
\end{figure}
Consider, for example, the decays $D^+\to\bar{\ka}^0\pi^+\to K^-\pi^+\pi^+$
and $D^+\to\bar{\ka}^0\ell^+\nu_\ell\to K^-\pi^+\ell^+\nu_\ell$. If
$\bar{\ka}^0$ is a quark anti-quark state ($\bar{\ka}^0=s\bar{d}$) the
above mentioned decays can procced through the diagrams in Fig.~1, from where
we see that one has two possible diagrams for the nonleptonic decay
(Figs.~1a and 1b), and only one diagram for the 
semileptonic decay (Fig.~1c). However,
if $\bar{\ka}^0$ is a four-quark state ($\bar{\ka}^0=us\bar{u}\bar{d}$),
from Fig.~2 we see that there are four possible diagrams for the
nonleptonic decay, while there is still only one diagram for the semileptonic
decay. Therefore, in the four-quark scenario the probability to form
a scalar meson from the decay of charmed mesons,
is much bigger for the nonleptonic decay
as compared with the semileptonic decay. A trend that seems be 
followed  by the experimental data \cite{ig}.
\begin{figure}[h] 
\label{fig2}
%\leavevmode
\centerline{\epsfig{figure=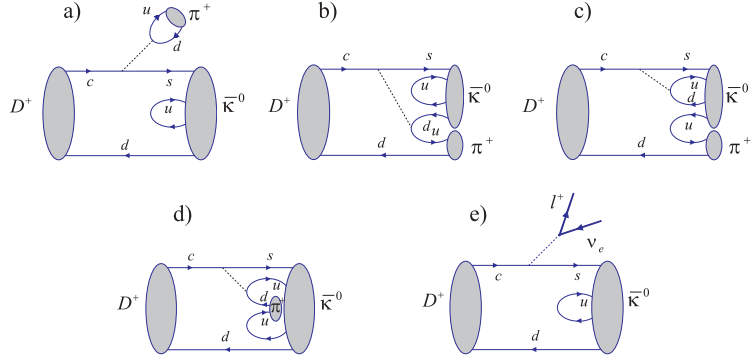,width=10cm,angle=0}}
\caption{The $D^+\to\bar{\ka}^0$ transition where $\bar{\ka}^0$ is described
as a $us\bar{u}\bar{d}$ state. a), b), c) and d) nonleptonic decay; 
e) semileptonic decay.}
\end{figure}

In this work we use the  method of QCD  sum rules (QCDSR) 
\cite{svz}
to study the two-point function and  the three-point function,
associated with the meson decay constant and hadronic coupling
contants of the scalar mesons, considered as four-quark states. The
first evaluation of the $f_0(980)$ as a four-quark state in the QCDSR formalism
was done in \cite{lapa} for the meson-vaccum decay constant, and in 
\cite{nari} for the hadronic coupling contants (for a review see
\cite{nari2}). We extend these works by considering all the scalar mesons
in the nonet and by considering different currents.

%\section{Two-point functions}

We follow ref.~\cite{mppr} and consider that the lowest lying scalar mesons
are $S$-wave bound states of a diquark-antidiquark pair. As suggested in 
ref.~\cite{jawil} the diquark is taken to be a spin zero colour anti-triplet, 
flavour
anti-triplet. Therefore, the $(q)^2(\bar{q})^2$ states make a flavour
$SU(3)$ nonet. The corresponding interpolating fields are:
\beqa
j_\si&=&\epsilon_{abc}\epsilon_{dec}(u_a^TC\gamma_5d_b)(\bar{u}_d\gamma_5C
\bar{d}_e^T),
\nn\\
j_{f_0}&=&{\epsilon_{abc}\epsilon_{dec}\over\sqrt{2}}\left[(u_a^TC\gamma_5s_b)
(\bar{u}_d\gamma_5C\bar{s}_e^T)+u\leftrightarrow d\right],
\nn\\
j_{a_0}&=&{\epsilon_{abc}\epsilon_{dec}\over\sqrt{2}}\left[(u_a^TC\gamma_5s_b)
(\bar{u}_d\gamma_5C\bar{s}_e^T)-u\leftrightarrow d\right],
\nn\\
j_\ka&=&\epsilon_{abc}\epsilon_{dec}(u_a^TC\gamma_5d_b)(\bar{q}_d\gamma_5C
\bar{s}_e^T),\;\;\;\bar{q}=\bar{u},\bar{d},
\label{int}
\enqa
where $a,~b,~c,~...$ are colour indices and $C$ is the charge conjugation
matrix. The other members of the nonet are easily constructed.

The coupling of the scalar meson $S$, to the scalar current $j_S$,  can be
parametrized in terms of the meson decay constant $f_S$ as \cite{lapa}:
\beq
\lag 0 | j_S|S\rag =\sqrt{2}f_Sm_S^4\;.
\lb{fs}
\enq
In order to compute this parameter by QCDSR, we consider the two-point
correlation function
\beq
\Pi(q)=i\int d^4x ~e^{iq.x}\lag 0 |T[j_S(x)j^\dagger_S(0)]|0\rag.
\lb{2po}
\enq
In the QCD side we work at leading order and consider condensates up to 
dimension six. We deal with the strange quark as a light one and consider
the diagrams up to order $m_s$. In the phenomenological side we consider
the usual pole plus continuum contribution. Therefore, we introduce
the continuum threshold parameter $s_0$ \cite{io1}. In the $SU(2)$ limit the
$f_0$ and $a_0$ states are, of course, mass degenerate, and we get the same
decay constant for them. After doing a Borel tranform the two-point sum rules
are given by:
\beqa
2f_\si^2m_\si^8e^{-m_\si^2/M^2}&=&{M^{10}E_4\over 2^95\pi^6}+{\lag g^2G^2\rag
M^6E_2\over2^{10}3\pi^6}+{\lag\bar{q}q\rag^2M^4E_1\over12\pi^2},
\nn\\
2f_\ka^2m_\ka^8e^{-m_\ka^2/M^2}&=&{M^{10}E_4\over 2^95\pi^6}-{m_s(2\lag
\bar{q}q\rag-\lag\bar{s}s\rag)\over2^63\pi^4}M^6E_2
+{\lag g^2G^2\rag M^6E_2\over2^{10}3\pi^6}
+{(\lag\bar{q}q\rag^2+\lag\bar{q}q\rag\lag\bar{s}s\rag)\over24\pi^2}M^4E_1
\nn\\
&+&{m_s\lag\bar{s}g\si.Gs\rag\over 2^83\pi^4}M^4E_1+
{m_s\lag\bar{q}g\si.Gq\rag\over 2^7\pi^4}\left({3\over2}-\ln(M^2/\Lambda^2)
\right) M^4E_1,
\nn\\
2f_{f_0}^2m_{f_0}^8e^{-m_{f_0}^2/M^2}&=&{M^{10}E_4\over 2^95\pi^6}-{m_s(2\lag
\bar{q}q\rag-\lag\bar{s}s\rag)\over2^53\pi^4}M^6E_2
+{\lag g^2G^2\rag M^6E_2\over2^{10}3\pi^6}
+{\lag\bar{q}q\rag\lag\bar{s}s\rag\over12\pi^2}M^4E_1
\nn\\
&+&{m_s\lag\bar{s}g\si.Gs\rag\over 2^73\pi^4}M^4E_1+
{m_s\lag\bar{q}g\si.Gq\rag\over 2^6\pi^4}\left({3\over2}-\ln(M^2/\Lambda^2)
\right) M^4E_1,
\lb{sr2}
\enqa
where
\beq
E_n\equiv 1-e^{-s_0/M^2}\sum_{k=0}^n\left(s_0\over M^2\right)^k{1\over k!}
\ ,
\label{con}
\enq
which accounts for the continuum contribution. 

\begin{figure}[h] \label{fig3}
%\leavevmode
\centerline{\epsfig{figure=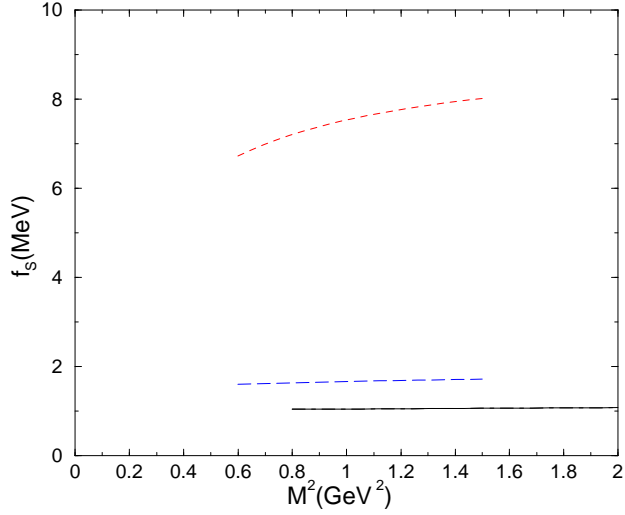,height=70mm}}
%,width=70mm,angle=0}}
\caption{The scalar meson decay constants as a function of the Borel mass.  
Solid: $f_{f_0}=f_{a_0}$; dashed: $f_\ka$;
dots: $f_\si$.} 
\end{figure} 
In the numerical analysis of the sum rules, the values used for the strange
quark mass and condensates are: $m_s=0.13\,\GeV$, $\lag\bar{q}q\rag=\,
-(0.23)^3\,\GeV^3$,
$\langle\overline{s}s\rangle\,=0.8\lag\bar{q}q\rag$,  
$\lag\bar{q}g\si.Gq\rag=m_0^2
\lag\bar{q}q\rag$ with $m_0^2=0.8\,\GeV^2$ and $\lag g^2G^2\rag=0.5~\GeV^4$.
We estimate the decay constants by using the experimental values for the
smalar meson masses \cite{E791}: $m_\si=0.5\,\GeV$, $m_\ka=0.8\,\GeV$,
$m_{f_0}=0.98\,\GeV$. For the continuum thresholds we use 
$s_0^\si=1.0\,\GeV^2$, $s_0^\ka=1.2\,\GeV^2$, $s_0^{f_0}=1.5\,\GeV^2$. 

From 
Fig. 1 we see that we get a very stable result, as a function of the Borel 
mass, for $f_{f_0}$ and $f_\ka$. In the case of $f_\si$ the stability is not
so good, but it is still acceptable. The problem with these sum rules, as
already noticed in ref.~\cite{lapa}, is that the continuum contribution is 
only smaller than the pole contibution for small values of the Borel mass
($M^2\leq 0.6\GeV^2$ for $\si$ and $\ka$ and $M^2\leq 0.8\GeV^2$ for $f_0$ 
and $a_0$). However, for these values of the Borel mass the perturbative 
contribution is smaller than the four-quark condensate contribution.
The perturbative contribution will get bigger than the four-quark condensate
contribution only for $M^2\sim1.5\GeV^2$ for $\si$ and $\ka$, and
$M^2\sim2.\GeV^2$ for $f_0$ and $a_0$.
Therefore, the Borel windows used is a compromise between these two
restrictions. Allowing a variation of $0.2\GeV^2$ in the continuum thresholds
we arrive at the following values for the decay constants:
\beq
f_\si=(7.5\pm1.0)\MeV,\;\;\;\;f_\ka=(1.6\pm0.3)\MeV,\;\;\;\;f_{f_0}=
(1.1\pm0.1)\MeV.
\enq

%\section{Three-point functions}

In order to study the scalar-pseudoscalar-pseudoscalar vertex associated with
the $\si\rightarrow\pi^+\pi^-$, $\ka\rightarrow K^+\pi^-$, 
$f_0(a_0)\rightarrow K^+K^-$ and $f_0\rightarrow\pi^+\pi^-$ decays, we 
consider the three-point function
\beq
T_{\mu\nu}(p,\pli,q)=\int d^4x~d^4y~e^{i.\pli.x}~e^{iq.y}\lag0|T\{
j_{5\mu}^{P_1}(x)j_{5\nu}^{P_2}(y)j_S^\dagger(0)\}|0\rag,
\lb{3point}
\enq
where $p=\pli+q$, $S$ denotes the scalar meson and $P_1,~P_2$ are 
the two pseudoscalar mesons in the vertex, for which we use the axial currents:
\beq
j_{5\mu}^{\pi^+}=\bar{d}_a\gamma_\mu\gamma_5u_a,\,\;\;\;j_{5\mu}^{\pi^-}=
\bar{u}_a\gamma_\mu\gamma_5d_a,\,\;\;\;j_{5\mu}^{K^+}=
\bar{s}_a\gamma_\mu\gamma_5u_a,\,\;\;\;j_{5\mu}^{K^-}=
\bar{u}_a\gamma_\mu\gamma_5s_a.
\lb{pseu}
\enq

In order to evaluate the phenomenological side
we insert intermediate states for $P_1$, $P_2$ and $S$, and we use the 
definitions 
in Eqs.~(\ref{fs}) and (\ref{fp}) bellow:
\beq
\lag 0 | j_{5\mu}^{P_i}|P_i(p)\rag =ip_\mu F_{P_i}\;.
\lb{fp}
\enq
We obtain the following relation
\beqa
&&T_{\mu\nu}^{phen} (p,\pli,q)={\sqrt{2}F_{P_1}F_{P_2}m_S^4f_S
\over (m_{S}^2-p^2)(m_{P_1}^2-{\pli}^2)(m_{P_2}^2-q^2)}~g_{SP_1P_2}~\pli_\mu 
q_\nu
+\mbox{ contributions of higher resonances}\;,
\lb{phen}
\enqa
where the coupling constant $g_{SP_1P_2}$ is defined by the matrix element
\beq
\lag P_1(\pli)P_2(q)|S(p)\rag=g_{SP_1P_2}.
\enq

Here we follow refs.~\cite{nari,nari2} and work at the pion pole, as suggested
in \cite{rry} for the nucleon-pion coupling constant.
This method was also applied to the nucleon-kaon-hyperon coupling 
\cite{ccl,brann}, to the $D^*-D-\pi$ coupling \cite{nnbcs} and to the
$J/\psi-\pi$ cross section \cite{nnmk}. It consists in neglecting the pion mass
in the denominator of Eq.~(\ref{phen}) and working at $q^2=0$. In the QCD side
one singles out the leading terms in the operator product expansion of 
Eq.(\ref{3point}) that mach the $1/q^2$ term. Up to dimension six only
the diagrams proportional to the quark condensate times $m_s$ and the 
four-quark condensate contribute.
Making a single Borel tranform to both $-p^2=-{\pli}^2\rightarrow M^2$ we get:
\beqa
g_{\si\pi^+\pi^-}{\sqrt{2}F_\pi^2f_\si m_\si^4\over m_\si^2-m_\pi^2}
(e^{-m_\pi^2/M^2}-e^{-m_\si^2/M^2})&=&{2\over3}\lag\bar{q}q\rag^2,
\nn\\
g_{\ka K^+\pi^-}{\sqrt{2}F_\pi F_Kf_\ka m_\ka^4\over m_\ka^2-m_K^2}
(e^{-m_K^2/M^2}-e^{-m_\ka^2/M^2})&=&{\lag\bar{q}q\rag\over3}(\lag\bar{q}q\rag
+\lag\bar{s}s\rag)-{m_s\over8\pi^2}\lag\bar{q}q\rag M^2\left(1-e^{-s_0^\ka/M^2}
\right),
\nn\\
g_{f_0 K^+K^-}{\sqrt{2}F_K^2f_{f_0} m_{f_0}^4\over m_{f_0}^2-m_K^2}
(e^{-m_K^2/M^2}-e^{-m_{f_0}^2/M^2})&=&{1\over6\sqrt{2}}(\lag\bar{q}q\rag^2
+\lag\bar{s}s\rag^2+\lag\bar{q}q\rag\lag\bar{s}s\rag)+
\nn\\
&-&{m_s\over16\sqrt{2}\pi^2}
\left(\lag\bar{q}q\rag+{2\over3}\lag\bar{s}s\rag\right) M^2
\left(1-e^{-s_0^\ka/M^2}
\right).
\lb{sr3}
\enqa

As said in the introduction, the $f_0\to\pi^+\pi^-$ decay is mediated by one 
gluon exchange. Therefore, the first diagram proportional to $1/q^2$
(at leading order)
is the eight dimension condensate proportional to the mixed condensate times
the quark condensate. Therefore, working up to dimension eight in this case
we get
\beq
g_{f_0\pi^+\pi^-}{\sqrt{2}F_\pi^2f_{f_0} m_{f_0}^4\over m_{f_0}^2-m_\pi^2}
(e^{-m_\pi^2/M^2}-e^{-m_{f_0}^2/M^2})={m_0^2\lag\bar{q}q\rag
\lag\bar{s}s\rag\over12\sqrt{2}M^2}.
\lb{sr32}
\enq

The problem of doing a single Borel transformation is the fact that terms
associated with the pole-continuum transitions are not suppressed \cite{io2}.
In ref.~\cite{io2} it was explicitly shown that the pole-continuum transitions
have a different behaviour, as a function of the Borel mass, as compared with
the double pole contribution: it grows with $M^2$. Therefore, the 
pole-continuum contribution can be taken into account through the introduction
of a parameter $A$ in the phenomenological side of the sum rules in 
Eqs.~(\ref{sr3}), (\ref{sr32}), by making the substitution
$g_{SP_1P_2}\rightarrow g_{SP_1P_2}+AM^2$ \cite{brann,nnbcs,nnmk}.

\begin{figure}[h] \label{fig4}
%\leavevmode
\centerline{\epsfig{figure=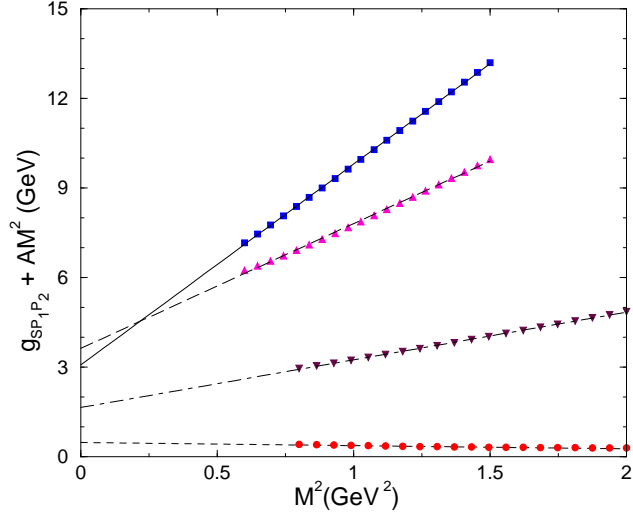,height=70mm}}
%,width=70mm,angle=0}}
\caption{The QCDSR results for the hadronic coupling constants
$g_{\si\pi^+\pi^-}$ (squares), $g_{\ka K^+\pi^-}$ (triangles up),
$g_{f_0 K^+K^-}$ (triangles down) and $g_{f_0\pi^+\pi^-}$ (circles). The
solid, dashed, dot-dashed and dotted lines give the linear fit to the
QCDSR results respectively.}
\end{figure} 

Using $F_\pi=\sqrt{2}~93\MeV$, $F_K=160\MeV$, $m_\pi=137\MeV$, $m_K=490\MeV$
and the scalar decay constant given by the sum rules in Eq.~(\ref{sr2})
we show, in Fig.~4 the QCDSR results for the hadronic coupling constants.
We see that, in the Borel range used for the two-point functions, the QCDSR
results do have  a linear behaviour as a function of the Borel mass. Fitting
the QCDSR results by a linear form: $g_{SP_1P_2}+AM^2$ (which is also shown
in Fig.~4), the hadronic couplings can be obtained by extrapolating the fit 
to $M^2=0$. 
In the  limits of the continuum thresholds discussed
above we obtain: 
\beqa
g_{\si\pi^+\pi^-}=(3.1\pm0.5)~\GeV,\;\;\;\;\;g_{\ka K^+\pi^-}=(3.6\pm0.3)~
\GeV,
\nn\\
g_{f_0 K^+K^-}=g_{a_0 K^+K^-}=(1.6\pm0.1)\GeV,\;\;\;\;\;g_{ f_0\pi^+\pi^-}
=(0.47\pm0.05)~\GeV.
\lb{coup}
\enqa

%\section{Deacy widths and conclusions}

The decay width of $S\rightarrow P_1P_2$ is given in terms of the hadronic
coupling $g_{SP_1P_2}$ as: 
\beq
\Gamma(S\rightarrow P_1P_2)={1\over 16\pi m_S^3}g_{SP_1P_2}^2\sqrt{\la(
m_S^2,m_{P_1}^2,m_{P_2}^2)},
\lb{decay}
\enq
where $\la(m_S^2,m_{P_1}^2,m_{P_2}^2)=m_S^4+m_{P_1}^4+m_{P_2}^4-2m_S^2
m_{P_1}^2-2m_S^2m_{P_2}^2-2m_{P_1}^2m_{P_2}^2$.

The experimental total decay width is related with a particular decay 
mode through:
\beqa
\Gamma(S\rightarrow\pi \pi)&=&{3\over2}\Gamma(S \rightarrow\pi^+ \pi^-),\;\;
\mbox{ for }S=\si \mbox{ or } f_0,
\nn\\
\Gamma(\ka\rightarrow K \pi)&=&{3\over2}\Gamma(\ka\rightarrow K^+ \pi^-).
\lb{rel}
\enqa
Therefore, using the experimental results: $\Gamma(f_0\rightarrow\pi \pi)=
40~-~100~\MeV$ \cite{PDG}, $\Gamma(\si\rightarrow\pi \pi)=(338\pm48)~\MeV$
\cite{E791} and $\Gamma(\ka\rightarrow K \pi)=(410\pm58)~\MeV$ \cite{E791}
in Eqs.~(\ref{decay}) and (\ref{rel}) above we get
\beq
g_{\si\pi^+\pi^-}^{exp}=(2.6\pm0.2)~\GeV,\;\;\;
g_{\ka K^+\pi^-}^{exp}=(4.5\pm0.4)~\GeV,\;\;\;
g_{ f_0\pi^+\pi^-}^{exp}=(1.6\pm0.8)~\GeV.
\lb{exp}
\enq

Comparing Eqs.~(\ref{coup}) and (\ref{exp}) we see that, although not exactly 
in between the experimental error bars, the hadronic couplings determined
from the QCDSR calculation are
consistent with existing experimental data. The biggest discrepancy is for
$g_{f_0\pi^+\pi^-}$ and this can be understood since, probably in this case,
$\alpha_s$ corrections could play an important role.
In the case of the decay
$f_0(a_0)\rightarrow K^+K^-$, the coupling can not be experimentally measured
due to the unavailable phase space.

We have presented a QCD sum rule study of the scalar mesons considered as
diquark-antidiquark states.
 We have evaluated the mesons decay contants and the hadronic couplings
associated with the $\si\rightarrow\pi^+\pi^-$, $\ka\rightarrow K^+\pi^-$, 
$f_0(a_0)\rightarrow K^+K^-$ and $f_0\rightarrow\pi^+\pi^-$ decays, using
two-point and thre-point functions respectivelly. We found that
the couplings are
consistent with existing experimental data. Therefore, we consider this 
result as one more point in favor of the four-quark structure for the light 
scalar mesons.

\vspace{1cm}
 
\underline{Acknowledgements}: 
We would like to thank I. Bediaga  for fruitful discussions. 
This work has been supported by CNPq and FAPESP.

\end{document}